\documentclass[prl,twocolumn,superscriptaddress,showpacs,%
preprintnumbers,amsmath,amssymb]{revtex4}
\usepackage{amsmath}
\usepackage{graphicx}
\usepackage{dcolumn}

\begin{document}

\title{Electrical control of inter-layer excitons in van der Waals heterostructures}

\author{A. Chaves} \email{andrey@fisica.ufc.br}
\affiliation{Departamento de F\'isica, Universidade Federal do
Cear\'a, Caixa Postal 6030, Campus do Pici, 60455-900 Fortaleza,
Cear\'a, Brazil}
\author{J. G. Azadani}
\affiliation{Department of Electrical and Computer Engineering, University of Minnesota, Minneapolis, Minnesota 55455, USA}
\author{V. Ongun \"Oz\c{c}elik}
\affiliation{Andlinger Center for Energy and the Environment, Princeton University, Princeton, New Jersey 08544, USA}
\author{R. Grassi}
\author{T. Low}\email{tlow@umn.edu}
\affiliation{Department of Electrical and Computer Engineering, University of Minnesota, Minneapolis, Minnesota 55455, USA}

\date{ \today }

\begin{abstract}
We investigate excitons in stacked transition metal dichalcogenide (TMDC) layers under perpendicularly applied electric field, herein MoSe$_2$/WSe$_2$ van der Waals heterostructures. Band structures are obtained with density functional theory calculations, along with the electron and hole wave functions in conduction and valence bands, respectively. Although the type-II nature of the heterostructure leads to fully charge separated inter-layer excitons, charge carriers distribution among the layers is shown to be easily tunable by external field. Our results show that moderate values of electric field produce more evenly distributed wave functions along the heterostructure, thus enhancing both the inter-layer exciton binding energy and, most notably, its oscillator strength. 
\end{abstract}


\maketitle
\emph{Introduction} - Recent experimental and theoretical analysis of transition metal dichalcogenides (TMDCs) have demonstrated that these indirect gap materials, when exfoliated into their monolayer form, acquire a direct gap at the $K$-point of the first Brillouin zone edge. \cite{Mak, Ganatra, Sun, Zhang, Ellis, Chei, Kumar, Book} Across the different types of transition metal and chalcogen combinations, survey of their optical gap \cite{TonyReview, Ye, Zhu, Ugeda, Tongay, Aslan}, excitonic Rydberg spectra \cite{Alexey, He}, trion and biexciton formation \cite{Sie, Ross, Wang} provide evidence of strong electron-hole interaction, due to the reduced dielectric screening in these systems. Van der Waals (vdW) heterostructures, through either vertical stacks or lateral attachment of different materials, present a promising avenue to engineering excitonic properties. Several experimental groups have reported advances in this direction, particularly on the investigation of photoluminescence (PL) in vertically stacked vdW bilayers. \cite{Miller, Rivera, Rigosi, Fang, Gong, Lin, Pak}

Recent efforts on vdW heterostructures have been focused on the pursuit for experimental evidence of spatially indirect inter-layer excitons (ILE), i.e. where the electron and hole forming the excitonic pair are confined at different materials. This situation is expected to occur in combinations of TMDCs with type-II band alignment \cite{Ongun}, i.e. both the conduction and valence band of one layer are at higher energy as compared to the same bands in the other layer. Such a spatially separated electron-hole pair should in general have very small electron-hole overlap and, consequently, longer recombination time, or excitons lifetime. This could potentially be of interest to energy storage and photodetector applications, where long exciton lifetimes are required. \cite{Lee} However, ILE has been proven to be elusive in experiments, since its very low oscillator strength forbids an inter-layer valence-to-conduction band transition to occur in light absorption measurements.

Here, we investigate electronic and excitonic properties of MoSe$_2$/WSe$_2$ heterostructures under an applied perpendicular electric field. Our results reveal strong electrical control of excitons binding energies and its oscillator strength. The former and latter would dictate the PL spectral resonance position and intensity respectively. Electronic band structures are obtained using density functional theory (DFT) calculations, whereas excitonic states are described within the Wannier-Mott approach.\cite{Tim, Olsen} In the absence of electric field, quasi-particle states around $K$-point are shown to be fully charge separated between the layers, due to the type-II bands mismatch, thus forming a vertical intrinsic electric dipole. As the field is applied opposite to the dipole direction, it pushes single electron and hole states towards each other across the bilayer, forcing the dipole to flip. At the crossover electric field, the dipole moment is minimized. In this case, the inter-layer exciton oscillator strength and binding energies are shown to attain maximum values, which suggests e.g. the possibility of observing a higher ILE peak and a visible blueshift of its position in PL experiments. The application of such an intermediate field can be achieved either by gating or doping one of the layers. The possibility of having a reciprocal space indirect inter-layer exciton (between $\Gamma$ and $K$ points) is also discussed.

\begin{figure*}[!t]
\centerline{\includegraphics[width=0.8\linewidth]{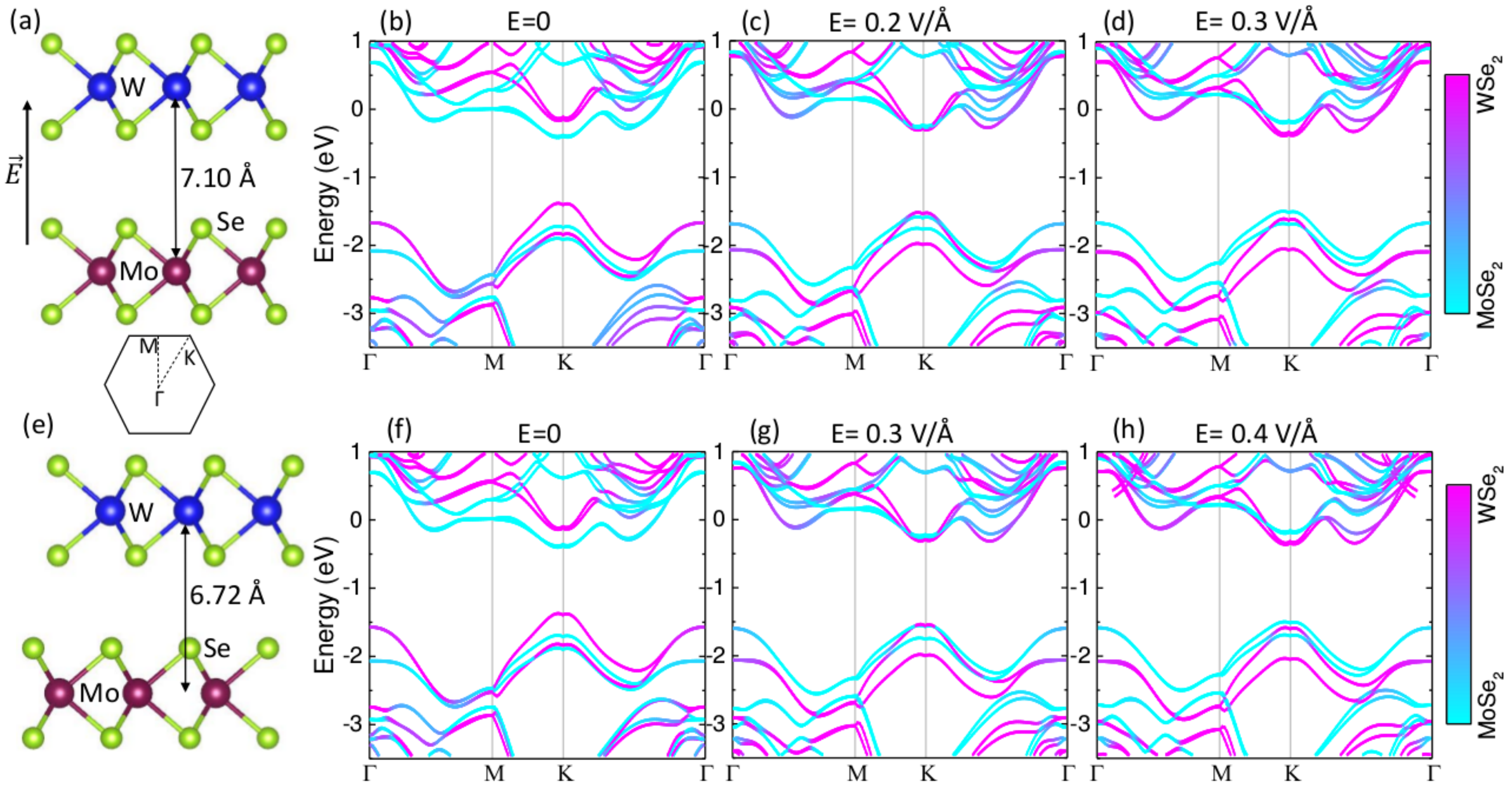}}
\caption{(color online) Side view of the crystal structure of MoSe$_2$/WSe$_2$ heterostructure with (a) AA and (e) AA' staking orientation. Inset: first Brillouin zone and its high symmetry points. In-plane lattice constant $a$ is 3.33\AA\,, while interlayer distances are 7.10\AA\, and 6.72\AA\, for the AA and AA' cases, respectively. The vertical arrow illustrates the positive direction of applied electric field $\vec E$. Band structures of AA stacked MoSe$_2$/WSe$_2$ heterostructure (b) without applied electric field, and with applied fields (c) $E = 0.2$ V/\AA\, and (d) $E = 0.3$ V/\AA\, are shown. (f-h) The same as (b-d), but for AA' stacking and $E = 0.3$ V/\AA\, and $E = 0.4$ V/\AA\,. The color scale indicates the relative wave function projections to the MoSe$_2$ (blue) and WSe$_2$ (pink) layers.}
\label{fig:bands} 
\end{figure*}

\emph{Theoretical model} - In order to obtain the band structure and electron/hole distributions for the vdW heterostructure, DFT calculations were performed using the Vienna \textit{ab initio} simulation package (VASP).\cite{VASP} We used projector-augmented wave (PAW)\cite{PAW} potentials and approximated the exchange-correlation potential with the Perdew-Burke-Ernzerhof (PBE)\cite{PBE} functionals. Van der Waals corrections \cite{Grimmes} and spin-orbit coupling were included in the calculations. During the ionic relaxation, the Brillouin zone of the unit cell was sampled by a grid of $(17\times 17\times 3)$ k-points. A plane-wave basis set with energy cutoff of 400 eV was used in all calculations. A large out-of-plane supercell size of 36 \AA{} was chosen to ensure no interaction between adjacent supercells. While applying external electric field, dipole corrections were applied in order to remove spurious dipole interactions.  \cite{Makov}

In general, TMDCs bilayers are shown to be indirect gap semiconductors. \cite{Ganatra, Mak} However, vdW heterostructures with AA stacking, instead of the natural AB form, exhibit direct gap at K. \cite{Lin} Electronically, these two layers are largely decoupled, with superposition of their monolayers bandstructure. Alternatively, AA' stacked bilayers, namely, with transition metal atoms of one layer superposing chalcogen atoms of the other layer, also share the same qualitative features. In our calculations, the geometry of AA and AA' stacked MoSe$_2$/WSe$_2$ supercells was first optimized at zero electric field, leading to the crystal structures shown in Figs. \ref{fig:bands}(a) and \ref{fig:bands}(e). For any other value of perpendicular electric field, we computed the optimum interlayer distance, by allowing the ions to relax in the out-of-plane direction until the atomic forces on the unit cell were less than 0.01 eV/\AA\,. The charge density was computed by refining the solution to the Kohn-Sham equations with an energy convergence value of $10^{-5}$ eV. While the above calculations were performed in the absence of spin-orbit coupling, a final run including spin-orbit coupling was used to compute wave functions and energy bands. Energies were calculated along the symmetry lines of the Brillouin zone shown in the inset of Fig. \ref{fig:bands}(a). Energies in all DFT results throughout the paper are shown using vacuum energy at zero electric field as reference.

\begin{figure}[!t]
\centerline{\includegraphics[width=\linewidth]{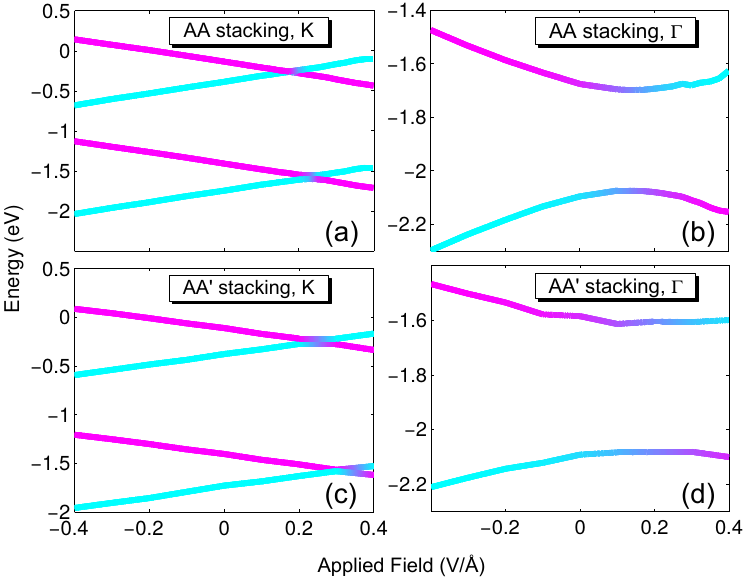}}
\caption{(color online) (a) Two lowest conduction band states and highest valence band states (VBM) for AA stacked MoSe$_2$/WSe$_2$ heterostructure under perpendicularly applied electric field at K point. (b) Valence band extrema at $\Gamma$ point as a function of the electric field. The same results, but for AA' stacking, are shown in (c) and (d), respectively. Color coding is the same as in Fig. \ref{fig:bands}, thus indicating the relative wave function projections to the MoSe$_2$ (blue) and WSe$_2$ (pink) layers.}
\label{fig:band_edges}
\end{figure}

From the pseudo-wave function $\Psi_{n,\vec{k}}(\vec{r}) = [ \psi^{\uparrow}_{n,\vec{k}}(\vec{r}), \psi^{\downarrow}_{n,\vec{k}}(\vec{r})]^T$ of a specific band $n$ and $\vec{k}$-point (in what follows, we choose either $\vec{k}=\mathrm{K}$ or $\vec{k}=\Gamma$), we extracted the probability function projection \cite{Proj} to the bottom layer (MoSe$_2$) as
\begin{equation}
\Phi^{(n)}_1 = \int_{z_{\mathrm{min}}}^{z_0} dz \int_{\Omega_\mathrm{2D}} dx \, dy \left[ \left| \psi^{\uparrow}_{n,\vec{k}}(\vec{r}) \right|^2 + \left| \psi^{\downarrow}_{n,\vec{k}}(\vec{r}) \right|^2\right]
\label{eq:projection}
\end{equation}
and similarly (but integrating from $z_0$ to $z_{max}$) for the top layer (WSe$_2$)
\begin{equation}
\Phi^{(n)}_2 = \int_{z_0}^{z_{\mathrm{max}}} dz \int_{\Omega_\mathrm{2D}} dx \, dy \left[ \left| \psi^{\uparrow}_{n,\vec{k}}(\vec{r}) \right|^2 + \left| \psi^{\downarrow}_{n,\vec{k}}(\vec{r}) \right|^2\right],
\label{eq:projection2}
\end{equation}
where $\Omega_\mathrm{2D}$ stands for the two-dimensional Wigner-Seitz cell, $z_{\mathrm{min}}$ and $z_{\mathrm{max}}$ are the supercell lower and upper boundaries in the out-of-plane direction, respectively, and $z_0$ is the mid-point position between the two layers. 

We define the exciton wave function, within the Wannier-Mott model, as $\Psi(r_e,r_h,i,j) = \psi_e(r_e)\psi_h(r_h)\phi_{ij}$, for electron and hole in the $i$-th and $j$-th layers, respectively. The exciton Hamiltonian in this basis reads 
\begin{equation} \label{eq.Ham}
H_{exc} = \sum_{i,j = 1}^{2}\left[\frac{1}{\mu^{ij}}\nabla^2_{\parallel} + V_{ij}(r_e-r_h)\right],
\end{equation}
where $i,j = 1(2)$ stands for the bottom (top) layer and $\mu_{ij}$ is the reduced effective mass for an electron (hole) in the $i(j)$-th layer. The exciton binding energy can then be obtained from $\langle \Psi | H_{exc} |\Psi \rangle$, which leads to
\begin{equation}
E_{exc} = E_{11} + E_{12} + E_{21} + E_{22},
\end{equation}
where $E_{ij} = |\phi_{ij}|^2E^{ij}_{\parallel}$, and $E^{ij}_{\parallel}$ is the energy obtained by numerically solving Schr\"odinger equation for the in-plane motion considering $V_{ij}(r_e-r_h)$ as the electron-hole interaction potential. The latter is obtained from the quantum electrostatic heterostructure model, \cite{Thygesen} assuming an electron in the $i$-th layer and a hole in the $j$-th layer. $|\phi_{ij}|^2$ is calculated by first integrating the square modulus of the DFT obtained electron (hole) wave function in the surroundings of the $i$($j$)-th layer (i.e. around $z_{min} < z < z_0$ and $z_{0} < z < z_{max}$ for bottom and top layers, respectively), such as in Eqs. (\ref{eq:projection}) and (\ref{eq:projection2}), in order to obtain the contribution from each layer to the overall probability density distribution. Multiplying the results for electrons and holes yields the probability of simultaneously having an electron in $i$ and a hole in $j$, namely $\phi_{ij} = \Phi^{(e)}_i\Phi^{(h)}_j$, where $e$ ($h$) stands for the lowest (highest) conduction (valence) band states. In this way, the electron-hole overlap is given the probability of having both quasi-particles at layer 1 \textit{or} layer 2, i.e. $|\phi_{11}|^2 + |\phi_{22}|^2$. 

\emph{Results} - DFT results in Figs. \ref{fig:bands}(b-d) and \ref{fig:bands}(f-h) show that the MoSe$_2$/WSe$_2$ heterostructure is a direct gap semiconductor at K point under all values of applied electric field investigated here, for both stacking configurations. Figures \ref{fig:bands}(b) and \ref{fig:bands}(f) show the band structure of the AA and AA' stacked heterostructures at zero electric field. In both cases, from the color scale, one verifies that the wave functions at the K point are well localized in either the MoSe$_2$ or WSe$_2$ layer. Hence, the combined band diagram of the heterostructure preserves the properties of the band structure of its constituent monolayers. This observation supports the idea of using the layer-localized states ($\phi_{ij}$) as a basis in which the Hamiltonian in Eq. (\ref{eq.Ham}) becomes diagonal. Otherwise, non-zero off-diagonal (hopping) terms would produce energy bands with layer-delocalized electron and hole wave functions instead.

\begin{figure}[!b]
\centerline{\includegraphics[width=\linewidth]{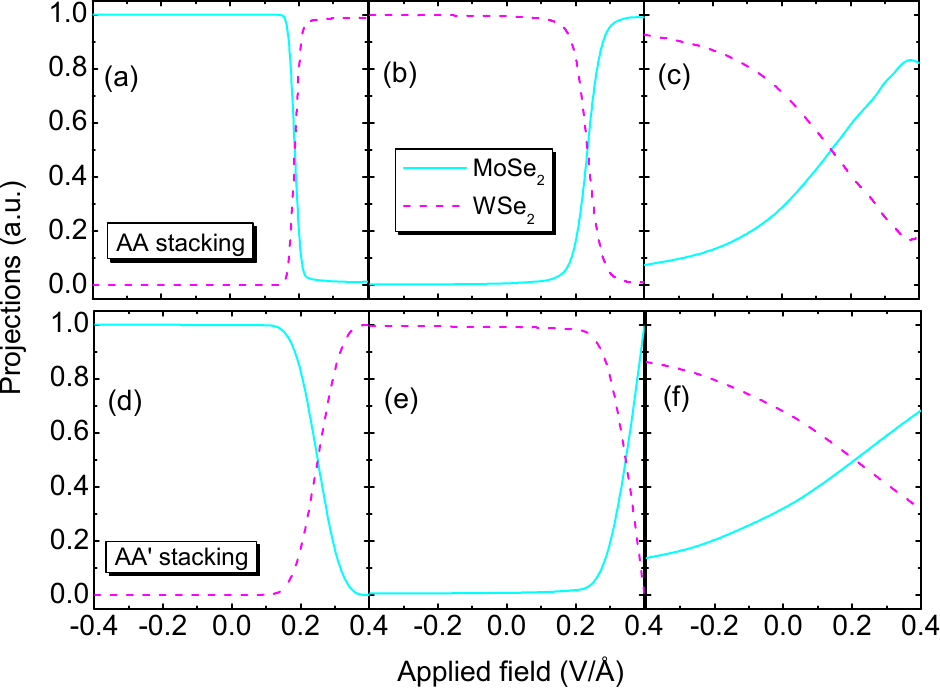}}
\caption{(color online) Normalized wave function projections on the MoSe$_2$ (solid) and WSe$_2$ (dashed) layers, calculated according to (\ref{eq:projection}) for (a) CBM and (b) VBM at K point, as well as for (c) VBM at $\Gamma$ point, assuming a AA stacked vdW heterostructure. (d-f) The same results, but for the AA' stacking case.}
\label{fig:projections}
\end{figure}

When a positive electric field is applied (see Fig. \ref{fig:bands}(a)), the bands around K with prevalent MoSe$_2$ character shift to higher energies, while those with WSe$_2$ character shift to lower energies, as one would expect if the band structures were uncoupled (see Fig.~\ref{fig:bands}(c) and Fig.~\ref{fig:bands}(d) at 0.2 and 0.3 V/\AA\,, respectively). This eventually results in a band crossing, which occurs around 0.2 V/\AA\, for both conduction and valence bands at K. The same situation holds for AA' stacked heterostructures, as shown in Figs. \ref{fig:bands}(f-h), albeit at a higher crossover electric field of $\approx$ 0.3 V/\AA\, instead. A similar effect is also discussed in Ref. [\onlinecite{Kaxiras}].

The situation is more complicated, however, for the $\Gamma$ point, where the valence band states have highly mixed projections. Despite the fact that the valence band maximum is at K, the satellite valley at $\Gamma$ is very close in energy and might be experimentally accessible as well. In fact, a recent PL experiment suggests the observation of an indirect exciton composed of an electron at K and a hole at $\Gamma$ in a similar heterostructure, namely MoS$_2$/WSe$_2$. \cite{Jens} The hole wavefunction at the $\Gamma$ point is found to be naturally distributed across both layers, despite the type-II nature of the heterostructure, thus leading to a high electron-hole overlap and higher recombination rate, as we will discuss in greater detail further on.

The energies of the two lowest conduction bands and the two highest valence bands at K point are plotted as a function of electric field in Fig.~\ref{fig:band_edges}(a). It is clear that the band gap energy of the heterostructure can be modulated by the electric field, reaching its maximum value of 1.28 eV at 0.225 V/\AA{}. Valence band extrema at the $\Gamma$ point are plotted as a function of the field in Fig.~\ref{fig:band_edges}(b), where an important qualitative difference, as compared to the valence band at K, is evidenced: the bands crossing observed in K-point states for a field around 0.225 V/\AA{} is replaced by an \textit{anti}-crossing at $\approx$ 0.185 V/\AA\, for the valence $\Gamma$ states. This suggest the existence of mixed layer states, namely, holes states at $\Gamma$ are delocalized across the two layers, in vicinity of the observed anti-crossing. Similar qualitative features are observed also in AA' stacking case in Figs. \ref{fig:band_edges}(c) and \ref{fig:band_edges}(d). The intrinsic distribution of $\Gamma$ states in valence band over both layers is confirmed by the wave function calculations in what follows.

\begin{figure}[!t]
\centerline{\includegraphics[width=\linewidth]{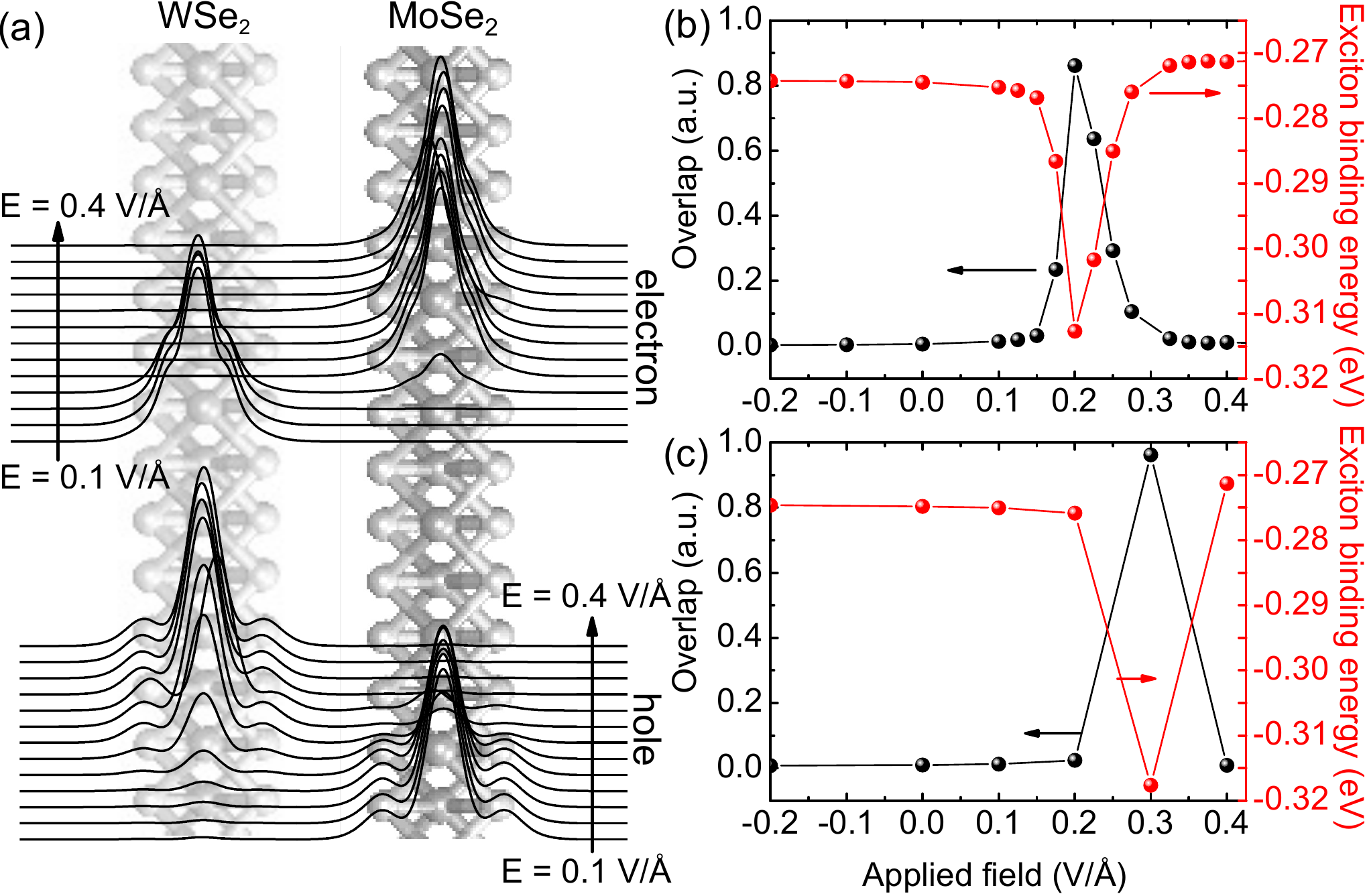}}
\caption{(Color online) (a) Electron and hole probability densities around the WSe$_2$ and MoSe$_2$ layers, for different values of applied electric field $E$, ranging from 0.1 to 0.4 V/\AA\, in steps of 0.025 V/\AA\,. Electron-hole overlaps (black, left scale) and exciton binding energies (red, right scale) as a function of the applied field, assuming (b) AA and (c) AA' stacking.} 
\label{fig:andrey}
\end{figure}

Wave function projections at K and $\Gamma$ points, calculated according to the definition in Eq. (\ref{eq:projection}), are shown in Fig. \ref{fig:projections} as a function of applied electric field. Figure \ref{fig:projections}(a) shows the projections of the lowest conduction band minimum (CBM) at K point, for AA stacking. For electric field values lower than 0.175 V/\AA\, (including the negative values), the wave function is almost completely localized in the first layer (MoSe$_2$). As the applied electric field increases, there is an abrupt shift to the other layer (WSe$_2$). 
For the valence band maximum (VBM) at K, shown in Fig. \ref{fig:projections}(b), this trend is reversed, as expected from the type-II band alignment. Besides, the transition is smoother for VBM, indicating a more pronounced band mixing at the band crossing. As for the valence band states at $\Gamma$ point, shown in Fig.~\ref{fig:projections}(c), the behavior is almost the same as for the hole states at K point, but there is an even smoother shift of wave functions from one layer to the other. We also note an intrinsic $\approx$65$\%$-35$\%$ distribution of this wave function between both layers, even in the absence of applied field. Similar results are also observed for the AA' stacking case in Figs. \ref{eq:projection}(d-f), although the crossover occurs at a higher electric field, in accordance to the bands crossing shown in Fig. \ref{fig:band_edges}(c) and \ref{fig:band_edges}(d).

The DFT obtained electron and hole probability densities in the AA stacking case are illustrated in Figure \ref{fig:andrey}(a), for different values of the perpendicularly applied electric field $E$. It is clear that for low fields, electrons (holes) are strongly localized in the WSe$_2$ (MoSe$_2$) layer, as previously discussed. As the applied field becomes more positive, charge carriers are pushed towards the opposite layer, and the localization of these particles is interchanged. For intermediate values of the applied field, however, the wave functions partially populate both layers and, consequently, a stronger electron-hole population overlap is observed. This is verified in Fig. \ref{fig:andrey}(b), for the AA stacked heterostructure, where this overlap reaches a maximum of $\approx$86$\%$ for $E \approx 0.2$ V/\AA\,, so that the modulus of the binding energy increases by  $\approx$ 45 meV. Despite the strong enhancement of the overlap, the increase in binding energy is small because the combination of binding energies $E_{ij}$ for each charge distribution case cannot be higher than the intra-layer values $E_{11} = 0.270$ eV (WSe$_2$) and $E_{22} = 0.320$ eV (MoSe$_2$). Inter-layer excitons have naturally lower binding energies $E_{12} = 0.274$ eV and $E_{12} = 0.271$ eV. The intensity (oscillator strength) of excitonic peaks (or its associated PL) is proportional to the square modulus of the exciton envelope wave function at $r = 0$, i.e. where electrons and holes are at the same point in space. Therefore, the electron-hole overlap is a measure of the exciton peak intensity and, consequently, the effects discussed here would manifest themselves in PL experiments as an inter-layer exciton peak whose spectral position and amplitude changes with external electric field. Results for the AA' stacked case are shown in Fig. \ref{fig:andrey}(c) and exhibit similar qualitative features, but with a 46 meV maximum variation of the exciton binding energy at $E \approx 0.3$ V/\AA\,, where the maximum overlap is $\approx$96$\%$, i.e. even higher than that of the AA stacked case.

\emph{Conclusion} - We have investigated the exciton states in MoSe$_2$/WSe$_2$ heterostructures in the presence of an electric field applied perpendicularly to the layers. Our results show that for almost any value of electric field, this heterostructure clearly exhibits type-II band offsets for states around the K-point of the Brillouin zone. This can be inferred when both electron and hole wave functions are strongly localized in separate layers. However, at the crossover electric fields of around 0.2 (0.3) V/\AA\,, for AA (AA') stacking, charge carriers distributions are flipped. In the close vicinity of this crossover electric field, the charge densities (especially the one for holes) are distributed across both layers. In this case, the electron-hole overlap and, consequently, exciton binding energy and oscillator strength can be significantly enhanced. 


The possibility of tuning the exciton binding energy and its PL emission intensity using an applied field opens an interesting avenue towards the search and enhancement of inter-layer charge-separated exciton states, whose experimental observation has proved to be elusive.

\acknowledgments  Discussions with J. Kunstman, D. R. Reichman, and C. W. Wong are gratefully acknowledged. AC has been financially supported by CNPq, through the PRONEX/FUNCAP and PQ programs. JA and TL acknowledge support from NSF ECCS-1542202. We acknowledge computational support  from the Minnesota Supercomputing Institute (MSI).

\end{document}